# PPV modelling of memristor-based oscillator

Bo Wang, Hanyu Wang, Miao Qi

In this letter, we propose for the first time a method of abstracting the PPV (Perturbation Projection Vector) characteristic of the up-to-date memristor-based oscillators. Inspired from biological oscillators and its characteristic named PRC (Phase Response Curve), we build a bridge between PRC and PPV. This relationship is verified rigorously using the transistor level simulation of Colpitts and ring oscillators, i.e., comparing the PPV converted from PRC and the PPV obtained from accurate PSS+PXF simulation. Then we apply this method to the PPV calculation of the memristor-based oscillator. By keeping the phase dynamics of the oscillator and dropping the details of voltage/current amplitude, the PPV modelling is highly efficient to describe the phase dynamics due to the oscillator coupling, and will be very suitable for the fast simulation of large scale oscillatory neural networks.

*Introduction:* The studies of oscillators are interdisciplinary [1-3], and the various models of oscillator have been proposed in the analysis of electronic clock circuits, biological neurons, and chemical reactions, etc. In micro/nano electronic domain, the oscillators not only provide precise clock signals, but also constitute ONN (oscillatory neural networks), in which the synchronization of the coupled oscillators allows for the realization of the tasks such as pattern recognition.

One of up-to-date and promising electronic oscillators is the memristor-based oscillator (Fig. 1) [4]. The inherent negative differential resistance (NDR) of the memristor allows compensating the energy loss and sustaining the stable oscillation. Its ultra-small chip area makes it very favourable to construct large scale neural network.

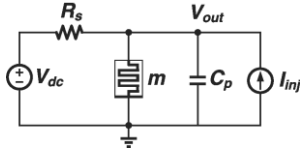

Fig.1 Memristor-based oscillator with impulse current injection $I_{inj}$

To describe the memristor, many physical and behavioural models have been proposed, and almost all of them are written in Verilog-A. The physical model often encounters the convergence problems due to its strong non-linearity. A more efficient and general behavioural model was proposed by Leon Chua based on unfolding theory, which uses the polynomial to approximate the highly nonlinearity of the memristor [4]. The oscillator can then be described by differential equations with coefficients determined by rigorous experiments, as shown below. $V_m$ and $x$ are respectively the voltage and the state variable of the memristor.

$$\frac{dv_m}{dt} = \frac{V_{dc} - v_m}{R_s C_p} - \frac{v_m}{C_p} \times \sum_{i=0}^{5} d_i x^i + \frac{I_{inj}\delta(t-t_1)}{C_p} \quad (I_{inj} = 0 ... if\ FreeRuning) \quad (1)$$

$$\frac{dx}{dt} = a_0 + a_1 x + b_2 v_m^2 + \sum_{i=1}^{5} c_{2i} v_m^2 x^i \quad (2)$$

However, these equations contain dozens of parameters ($a_i, b_i, c_i, d_i$) and show strong nonlinearity. While simulating the ONN with large number of oscillators, the simulation will become very slow (several hours or even days), because all the details of voltage / current and state variables are included during the simulation.

In this context we propose a much more efficient method: abstracting the PPV (phase characteristic) of the memristor-based oscillator, and using it to describe the behaviour of oscillators. This makes the modelling of oscillators very compact. The reason of doing this is the variation in oscillator's amplitude illustrates little influence on the oscillators and dies out with time, while the phase variation persists and eventually impacts the oscillator's behaviour [2]. Moreover, the PPV characteristic is general: it can be applied to describe any type of oscillator, regardless of its material or circuit structure [3].

*PPV:* For the electronic oscillators, when weak current is injected, the resulted deviation in phase is proportional to the quantities of the injected charge, and its impact depends on the timing of the injection. This can be described by linear time variant ISF (Impulse Sensitivity Function) function $\Gamma(t)$ [2]. In 2000, Demir [3] proposed a nonlinear time variant model called PPV ($\Gamma(t+\alpha)$), which further improved the ISF by considering the time shift of the ISF shortly after the impact of previous injected signal on the oscillator [5]. This allows PPV to analyse the continuous-time injection of oscillators, in addition to single injection. This makes it possible to analyse the coupled oscillators and the synchronization of the oscillatory neural networks.

Calculation of PPV: The PPV can be calculated using time integration but with poor precision [3]. The frequency domain Harmonic balance analysis is precise but complex. A precise and efficient method is proposed to use PSS/PXF of SpectreRF (Cadence) [6]. However, if the oscillator model is described in Verilog-A, which is the common case for memristor-based oscillator, PSS analysis fails to converge due to the hidden state in SpectreRF simulator.

*PRC:* On the other hand, in biological domain, people usually employ PRC to describe the phase evolution of the oscillator after the perturbation [1]. Although the types of the injected signals (magnetic force, light, etc.) could be different and diverse, the essential idea is same, i.e., to describe the impact of injection on the time/phase shift of the periodical behaviour. First proposed in 1948, and then studied by Kuramoto, Malkin and Winfree and Izhikevich [1], this method is popular and widely accepted.

Calculation of PRC: Kuramoto, Winfree and Malkin proposed respectively the formula of PRC [1]. One method is to inject an impulse signal to the oscillator, and then measure the phase shift of the oscillator due to the impulse after the re-stabilization. Another method proposed by Malkin is to use the backward integration to find the Jacobian matrix. Both are efficient; here we adopt the first method for its simplicity.

*Inspiration from PRC and PPV:* PPV (in rad/Ampere) is independent of the input under weak injection, which is preferable in circuit analysis. The analysis with PPV is rigorous and can be used to describe the coupling and synchronization of oscillator array [3]. However, the accurate calculation of PPV depends on the convergence of PSS or harmonic balance [6], as mentioned above.

PRC is easy to get, but it is specific to the inputs: different inputs produce different output phase shifts hence different PRC (in rad). If we can find a way to convert PRC to PPV, then we can benefit from the powerful strength of PPV mentioned above.

Suppose that we simulate respectively the PPV (using PSS+PXF) and PRC (using transient analysis) of the same electronic oscillator at the transistor level, we can find that their shapes are similar despite of the differences in amplitude and phase shift. Thus it gives a hint that it could be possible to find a transformation relation between them. This relation helps to bridge the gap between PPV and PRC. If encountering the difficulties in calculation of PPV for one oscillator (in our case the memristor-based oscillator), one can rely on the calculation of the PRC and then transform it to PPV according to this relationship.

*Conversion from PRC to PPV:* When the oscillator is free-running, its output can be represented as: $V_{out}(t) = f(\omega_0 t - \theta(t))$, where $f(...)$ is the periodic function.

If the injection signal is a small impulse $b(t)$ occurring at time $t_1$, then it will result in a phase shift to the oscillator whose quantity depends on the injection timing and strength. Here we use the zero crossing point as reference point for the calculation of phase shift. So we can define PRC the phase shift between the free-running oscillation $\theta_{fr}$ and the oscillation after injection $\theta_{inj}$:

$$PRC(t_1) = \theta_{inj}(t) - \theta_{fr}(t) \quad (3)$$

On the other hand, for PPV modelling, the time shift due to the current injection can be written as: [5]

$$\alpha(t) = \int_{-\infty}^{t} \Gamma(\tau + \alpha(\tau))b(\tau)d\tau \quad (4)$$

If the impulse $b(t)$ has a very brief width of $h$ and peak value of $b$, and occurs at $t_1$, then the time shift in eq.(4) can be approximated to:

$$\alpha(t) \approx \Gamma(t_1) \cdot h \cdot b \quad (5)$$

And its corresponding phase shift can be written as:

$$P(t_1) = \Gamma(t_1) \cdot h \cdot b \cdot \omega_0 \quad (6)$$

For the same signal injected to the same oscillators, the phase shift calculated by eq.(3) and eq.(6) should be identical, so we have:

$$PRC(t_1) = \theta_{inj}(t_1) - \theta_{fr}(t_1) = P(t_1) = \Gamma(t_1) \cdot h \cdot b \cdot \omega_0 \quad (7)$$

Hence PPV can be converted from PRC using following formula:

$$\Gamma(t_1) = \frac{PRC(t_1)}{h \cdot b \cdot \omega_0} \quad (8)$$

*Verification:* To verify this relationship between PPV and PRC, we calculate the PPV using the proposed PRC conversion method, and compare it with the PPV obtained by the accurate PSS/PXF simulation. We choose 2 electronic oscillators as examples [5]. One is Colpitts oscillator (Fig. 2), another is ring oscillator(Fig. 3); both are simulated at the transistor level to ensure the calculation accuracy.

First we calculate PPV using PSS+PXF simulation of two circuits. Note should be taken that:

(1) In PSS, choose *tstab* such that $\psi_1 = N * \pi/2$, where *tstab* is the stabilization time in PSS, $\psi_1$ the phase of fundamental component of the output voltage, and *N* any integer.
(2) After PSS+PXF simulation, if $\psi_1 - \phi \neq 0$ or $\psi_1 - \theta \neq \pi$, the phase of each harmonic component in PXF should be shifted by $-\phi$ (output phase of PXF at $+\Delta\omega$) or $\theta$ (output phase of PXF at $-\Delta\omega$).

Then we run transient simulation to get PRC. Within one period we select 100 different time points to inject the current impulse, and the simulation takes about 12 minutes.

Finally, we convert PRC to PPV using eq.(8), and superpose this PPV curve on the PPV curve calculated directly from PSS+PXF. As shown in Fig.2 and Fig.3, they all match very well. This confirms the correctness of our conversion method and paves the way to the PPV calculation of the memristor-based oscillators.

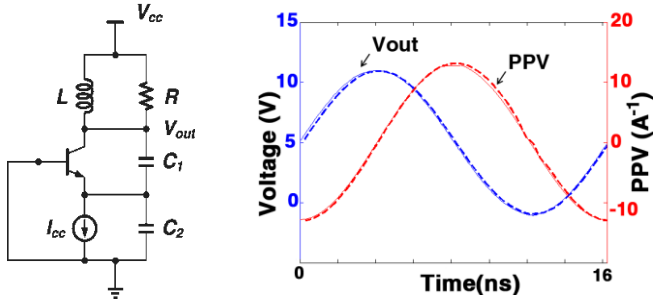

Fig. 2 (Left) Colpitts oscillator; (Right) PPV from PSS/PXF (solid line) vs. PPV converted from PRC (dashed line)

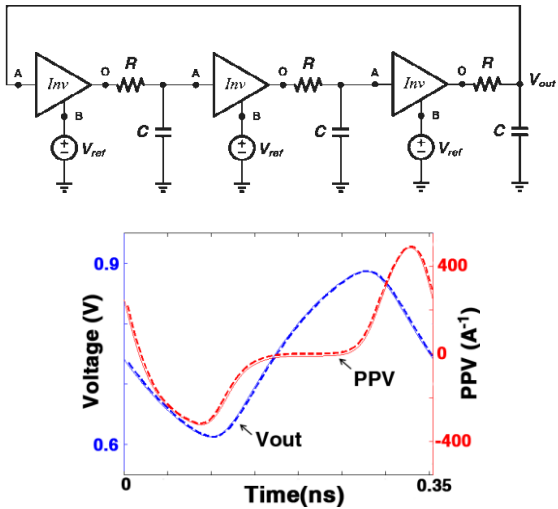

Fig.3 (Upper) Ring oscillator; (Lower) PPV from PSS/PXF(solid line) vs. PPV converted from PRC (dashed line)

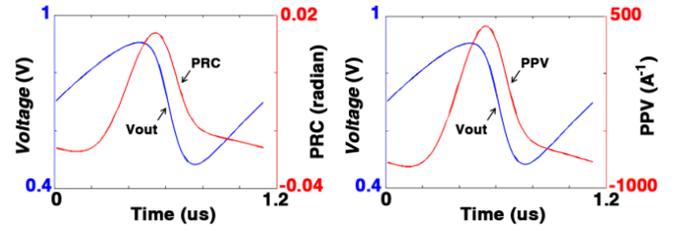

(a) PRC ($R_s$=1k ohms, $C_p$ = 3500pF)   (b) PPV corresponding to (a)

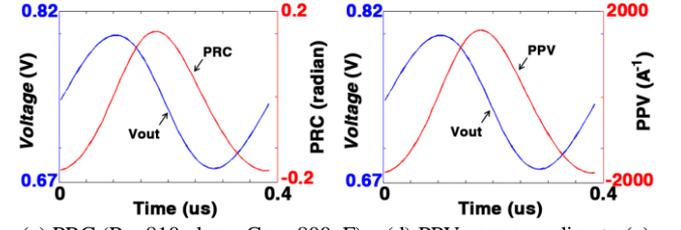

(c) PRC (Rs=810 ohms, Cp = 800pF)   (d) PPV corresponding to (c)

Fig. 4 PRC and PPV of memristor-based oscillator

*PPV of memristor-based oscillator:* Now we apply the proposed method to the memristor-based oscillator described by eq.(1-2). The equations are implemented in Verilog-A using Euler integration algorithm while adopting the same coefficients as those in [4]. We inject the impulse current into the output node of oscillator (shown in Fig.1), and measure the phase shift to get PRC (Fig. 4a). In mathematical equations, it is equivalent to add an impulse at the right side of differential equation, shown in last item of eq. (1): $I_{inj}\delta(t-t_1)/C_p$. Finally we convert PRC to PPV, as shown in Fig. 4b.

As a final experiment, we tune the serial resistor and parallel capacitor in the oscillator such that the output voltage approaches to the cosine wave (Fig.4 c, d). Interestingly we find that the PPV shape also changes to the similar form, but has a phase shift of 90 degree to the output voltage. This meets the expectations in reference [2] that a cosine output voltage corresponds to a sine PPV.

*Conclusion:* The relationship between PPV and PRC is identified and verified. Based on this, the PPV of memristor-based oscillator is determined. Focusing on the phase dynamics, the obtained PPV of the memristor-based oscillator allows for the accurate description of the phase coupling and synchronization. In future work it could be used for the efficient analysis of the large scale oscillator arrays.

*Acknowledgments:* Supported by R&D projects of Shenzhen city (JCYJ20150331102721193) and NSFC (61471011).

Bo Wang, Hanyu Wang, Miao Qi (*The Key Lab of IMS, School of ECE, Peking University, Shenzhen Graduate School*)
E-mail: wangbo@pkusz.edu.cn

**References**

1. E. Izhikevich, Dynamical Systems in Neuroscience: The Geometry of Excitability and Bursting. Cambridge, Mass.: The MIT Press, 2007.
2. A. Hajimiri and T. H. Lee, "A general theory of phase noise in electrical oscillators," *IEEE JSSC*, vol. 33, no. 2, pp. 179 –194, 1998.
3. A. Demir, A. Mehrotra, and J. Roychowdhury, "Phase Noise in Oscillators: A Unifying Theory and Numerical Methods for Characterization," *IEEE Trans. on CAS I*, vol. 47, no. 5, p. 655, 2000.
4. A. Ascoli, S. Slesazeck, H. Mahne, R. Tetzlaff, and T. Mikolajick, "Nonlinear Dynamics of a Locally-Active Memristor," *IEEE Trans. on CAS I*, vol. 62, no. 4, pp. 1165–1174, Apr. 2015.
5. P. Maffezzoni, "Analysis of oscillator injection locking through phase-domain impulse-response," *IEEE Trans. on CAS I*, vol. 55, no. 5, pp. 1297–1305, 2008.
6. S. Levantino and P. Maffezzoni, "Computing the Perturbation Projection Vector of Oscillators via Frequency Domain Analysis," *IEEE Trans. on CAD*, vol. 31, no. 10, pp. 1499–1507, Oct. 2012.